\documentstyle[twocolumn,aps,pra]{revtex}

\newcommand{\noun}[1]{\textsc{#1}}

\newcommand{\mathbold}[1]{\mbox{\boldmath $\bf#1$}}

\begin{document}

\tightenlines

\wideabs{

\title{Dissipative dynamics of a vortex state in a trapped Bose-condensed gas.}

\author{P.O. Fedichev\protect\( ^{1,2}\protect \) and G.V. Shlyapnikov\protect\( ^{1,2,3}\protect \)}

\address{\protect\( ^{1}\protect \)FOM Institute for Atomic and Molecular Physics,
Kruislaan 407, 1098 SJ Amsterdam, The Netherlands.}

\address{\protect\( ^{2}\protect \)Russian Research Center, Kurchatov Institute, Kurchatov
Square, 123182 Moscow, Russia.}

\address{\protect\( ^{3}\protect \)Laboratoire Kastler Brossel\protect\( ^{*}\protect \),
24 Rue Lhomond, F-75231 Paris Cedex 05, France}

\maketitle
\begin{abstract}
We discuss dissipative dynamics of a vortex state in a trapped Bose-condensed
gas at finite temperature and draw a scenario of decay of this state in a static
trap. The interaction of the vortex with the thermal cloud transfers energy
from the vortex to the cloud and induces the motion of the vortex core to the
border of the condensate. Once the vortex reaches the border, it immediately
decays through the creation of excitations. We calculate the characteristic
life-time of a vortex state and address the question of how the dissipative
dynamics of vortices can be studied experimentally. 

\end{abstract}
}

The recent successful experiments on Bose-Einstein condensation (BEC) in trapped
clouds of alkali atoms \cite{Cor95,Ket95,Hul95} have stimulated a great interest
in the field of ultra-cold gases \cite{Pitaevskii:review}. One of the goals
of ongoing studies is to investigate the nature of a superfluid phase transition
in ultra-cold gases and to make a link to more complicated quantum systems,
such as superfluid helium. Of particular interest is the relation between Bose-Einstein
condensation and superfluidity. However, being the most spectacular manifestation
of the phase transition in \( ^{4}{\rm He} \), superfluidity has not yet been
observed in trapped gases. A promising way of studying superfluidity in trapped
gases is the creation of quantum vortices, as quantization of circulation and
the related phenomenon of persistent currents are the most striking properties
of superfluids.

A widely discussed option of creating vortices in trapped gases assumes the
rotation of a slight asymmetry of a cylindrical trap after achieving BEC, or
cooling down the gas sample below the Bose-condensation temperature in an already
rotating trap\cite{Legget,Stringari:vortices2}. Another possibility is a rapid
quench of a gas sample near the critical temperature, which should lead to creation
of vortices even in a non-rotating trap \cite{Zurek}. It is worth mentioning
the ideas to create the vortex state in a Bose-condensed gas by optical means
\cite{Zoller:vorticesprod,olshanii:vortices}, and the idea to form vortex rings
in the regime of developed turbulence \cite{propTurb}. The spatial size of
the vortex core in the Thomas-Fermi regime is too small to be observed, and
for visualizing the vortex state it is suggested to switch off the trap and
let the cloud ballistically expand. Then the size of the vortex core will be
magnified approximately by the same factor as the size of the expanding condensate
\cite{YC:last}.

Similarly to the recently studied kink-wise condensates\cite{Zoller:vorticesprod,anglin:kink,gora:kink},
vortices are the examples of macroscopically excited Bose-condensed states.
In a non-rotating trap the vortex state has a higher energy than the ground-state
Bose condensate, i.e. the vortex is thermodynamically unstable \cite{Rokhsar,Fetter:dissipation,Fetter:recent}.
On the other hand, a quantum vortex with the lowest possible circulation (the
vortex ``charge'' equal to \( 1 \)), is dynamically stable (small perturbations
do not develop exponentially with time; see \cite{Bigelow,YC:last} and refs.
therein). Therefore, the vortex state can only decay in the presence of dissipative
processes. 

In this Letter we discuss dissipative dynamics of a vortex state in a trapped
Bose-condensed gas at finite temperatures and show how the interaction of the
vortex with the thermal cloud leads to decay of the vortex state in a static
trap. According to our scenario, the scattering of thermal excitations by a
vortex provides the energy transfer from the vortex to the thermal cloud and
induces the motion of the vortex core to the border of the condensate, where
the vortex decays by creating elementary excitations. We calculate the characteristic
life-time of the vortex state are discuss how the dissipative dynamics of vortices
can be studied experimentally. 

We first briefly outline the main features of the vortex behavior in a superfluid,
known from the studies of liquid helium. The motion of a vortex in a superfluid
of density \( \rho _{s} \) satisfies the Magnus law (see \cite{Donelly,Sonin,PAo:vortex}
and refs therein):
\begin{equation}
\label{eqofmotion}
\rho _{s}({\bf v}_{L}-{\bf v}_{S})\times \mathbold {\kappa }={\bf F}.
\end{equation}
Here \( {\bf v}_{L} \) is the velocity of the vortex line, and \( {\bf v}_{S} \)
the velocity of superfluid at the vortex line. The vector \( \mathbold {\kappa } \)
is parallel to the vortex line and is equal to the circulation carried by the
vortex. The force \( {\bf F} \) acting on the vortex originates from the mutual
friction between the normal component and the moving vortex line, and is usually
small. Assuming the absence of friction (\( {\bf F}=0 \)), the vortex moves
together with the superfluid component (\( {\bf v}_{L}={\bf v}_{S} \)). The
superfluid velocity \( {\bf v}_{S}({\bf r}) \) in the presence of a vortex
at the point \( {\bf r}_{0} \) satisfies the equations
\begin{equation}
\label{hydreq}
{\rm rot}{\bf v}_{S}=2\pi \mathbold {\kappa }\delta ({\bf r}-{\bf r}_{0});\; {\rm div}{\bf v}_{S}=0
\end{equation}
 and is related to the phase \( \phi  \) of the condensate wave-function as
\( {\bf v}_{S}=\mathbold {\nabla }\phi  \). This leads to quantization of the
circulation: \( \kappa =Z\hbar /m \) \cite{Onsager}, where \( Z \) is an
integer (the charge of the vortex) and \( m \) is the mass of the condensate
particle. Below we will consider vortex states with \( Z=1 \), which are dynamically
stable (\cite{Bigelow,YC:last} and refs. therein). 

Eqs.(\ref{hydreq}) are similar to the equations of the magnetostatic problem,
with the magnetic field replaced by the velocity \( {\bf v}_{S} \) and the
electric current replaced by \( \mathbold {\kappa } \). The velocity field
around an infinitely long straight vortex line is analogous to the magnetic
field of a straight current: 
\begin{equation}
\label{vel}
{\bf v}_{S}({\bf r})=[\mathbold {\kappa }\times {\bf r}]/r^{2}.
\end{equation}
The vortex itself can experience small oscillations of its filament, characterized
by the dispersion law \( |\omega (k)|=\kappa k^{2}\ln (1/ak)/2 \) \cite{SP2},
where \noun{\( k \)} is the wave vector of the oscillations, and \( a \)
the radius of the vortex core. In a weakly interacting Bose-condensed gas the
core radius is of order the healing length \( a=(\hbar ^{2}/m\mu )^{1/2} \),
where \( \mu  \) is the chemical potential.

We will see that the dissipative dynamics of a vortex state is insensitive to
the details of the density distribution in a gas. The spatial size of the Thomas-Fermi
condensate trapped in a harmonic potential of frequency \( \omega  \) is \( R=(2\mu /m\omega ^{2})^{1/2} \).
Therefore, for finding the superfluid velocity \( {\bf v}_{S} \) in this case,
we may consider a vortex in a spatially homogeneous condensate in a cylindrical
vessel of radius \( R \), with the vortex line parallel to the axis of the
cylinder. For the vortex line at distance \( x_{0} \) from the axis, the velocity
field can be found by using the ``reflection'' method\cite{Donelly}. In a
non-rotating trap, in order to compensate the normal component of the velocity
field (\ref{vel}) everywhere on the surface of the cylinder, we introduce a
fictious vortex with opposite circulation on the other side of the vessel wall,
i.e. at distance \( R^{2}/x_{0} \) from the cylinder axis. At the position
of the vortex the ``reflection'' induces the velocity
\begin{equation}
\label{veldrift}
{\bf v}_{S}=\frac{[\mathbold {\kappa }\times {\bf x}_{0}]}{R^{2}-x_{0}^{2}}.
\end{equation}
As \( v_{S}\approx v_{L} \), the vortex line will slowly drift around the axis
of the trap. A characteristic time responsible for the formation of the velocity
field (\ref{veldrift}) is \( \tau _{R}\sim R/c_{s} \), where \( c_{s}=\sqrt{\mu /m} \)
is the velocity of sound. Sufficiently far from the border of the Thomas-Fermi
condensate, i.e. outside the spatial region where \( R-x_{0}\ll R \), the drift
period is \( \tau _{dr}\sim x_{0}/v_{S}\sim R^{2}/\kappa  \) and greatly exceeds
the time \( \tau _{R} \) :
\[
\frac{\tau _{R}}{\tau _{dr}}\sim \frac{R}{c_{s}\tau _{dr}}\sim \frac{a}{R}\ll 1.\]
This means that we can neglect the retardation effects and, in particular, the
emission of phonons by the moving vortex. In other words, the ``cyclotron''
radiation is prohibited, since the wavelength \( c_{s}\tau _{dr} \) of sound
which would be emitted exceeds the size \( R \) of the condensate. 

According to the above mentioned magnetostatic analogy, in a non-rotating trap
the potential energy of the system (vortex plus its reflection) can be thought
as the energy of two counter flowing currents. Since the currents attract each
other, the energy is negative and decreases with displacing the vortex core
towards the wall. In other words, it is energetically favorable for the vortex
to move to the border of the vessel. Near the border the velocity of the vortex
exceeds the Landau critical velocity, and in a homogeneous superfluid the vortex
decays through the creation of phonons \cite{Donelly}. In a trapped gas the
condensate density strongly decreases near the border, and the vortex can decay
by emitting both collective and single-particle excitations. The motion of the
vortex towards the wall requires the presence of dissipation, as in the frictionless
approach the velocity of the vortex core coincides with the velocity (\ref{veldrift})
which does not contain a radial component. Thus, just the presence of dissipative
processes provides a decay of the vortex state (see \cite{Fetter:dissipation}
and related discussion \cite{gora:kink} of the stability of a kink state).

The dissipation originates from the scattering of elementary excitations by
the vortex and is related to the friction force \( {\bf F} \) in Eq.(\ref{eqofmotion}),
which is nothing else than the momentum transferred from the excitations to
the vortex per unit time. This force can be decomposed into longitudinal and
transverse components:
\begin{equation}
\label{forcegen}
{\bf F}=-D{\bf u}-D'{\bf u}\times \mathbold {\kappa }/\kappa ,
\end{equation}
where \( {\bf u}={\bf v}_{L}-{\bf v}_{n} \), \( {\bf v}_{n} \) is the velocity
of the normal component, and \( D,D' \) are longitudinal and transverse friction
coefficients, respectively. In a static trap \( {\bf u}={\bf v}_{L} \), as
the normal component is at rest (\( {\bf v}_{n}=0 \)). The friction force has
been investigated in relation to the attenuation of the second sound in superfluid
\( ^{4}{\rm He} \), where the most important is the transverse component \cite{Pitaevskii:friction,Fetter:friction,Iordanskii}
(see also \cite{Sonin} for review). For a straight infinite vortex line (parallel
to the \( z \)-axis), a general expression for the friction force in a homogeneous
superfluid is obtained in terms of the scattering amplitude \( f({\bf k},{\bf k}') \)\cite{Iordanskii}:
\[
{\bf F}=\left[ \int \frac{\partial n}{\partial E_{k}}\hbar ({\bf ku})\int ({\bf k}-{\bf k}')\frac{\delta (E_{k}-E_{k'})}{\delta (k_{z}-k_{z}')}\right. \]
\begin{equation}
\label{Fdef}
\left. |f({\bf k},{\bf k}')|^{2}\frac{d^{3}k}{(2\pi )^{3}}\frac{d^{3}k'}{(2\pi )^{3}}\right] -[{\bf u}\times \mathbold {\kappa }]\rho _{n}.
\end{equation}
Here \( \rho _{n} \) is the local mass density of the normal component, \( {\bf k},{\bf k}' \)
are the wave vectors of the incident and scattered excitations, \( n(E_{k})=(\exp (E_{k}/T)-1)^{-1} \)are
the Bose occupation numbers for the excitations, \( E_{k} \) is the excitation
energy, and \( T \) the gas temperature. Comparing the second terms of Eqs.
(\ref{forcegen}) and (\ref{Fdef}), one immediately arrives at the universal
expression for the transverse friction coefficient: \( D'=\kappa \rho _{n} \),
assuming that the first term of Eq.(\ref{Fdef}) does not contribute to \( D' \)
\cite{Sonin,Iordanskii}. 

We now turn to our analysis of the dissipative dynamics of the vortex state
in a non-rotating trap, related to the motion of the vortex core (line) to the
border of the condensate. This motion occurs on top of small oscillations of
the vortex filament and a slow drift (\ref{veldrift}) of the vortex core. The
radial component of the velocity of the vortex core is determined by the longitudinal
friction coefficient \( D \). For finding these quantities in dilute Bose-condensed
gases, the analysis of \cite{Pitaevskii:friction,Fetter:friction,Iordanskii,Sonin}
can only be used at very low temperatures (\( T\ll \mu  \)), where the number
of thermal excitations is very small and, hence, the longitudinal friction force
is extremely weak. 

The situation is drastically different in the temperature range \( T\agt \mu  \),
which is the most interesting for trapped Bose-condensed gases. We will consider
the limit \( T\gg \mu  \) and first analyze how the vortex scatters excitations
with energies \( E_{k}\agt \mu  \). These excitations are single particles,
and their De Broglie wave length is much smaller than the spatial size \( R \)
of the condensate. The most important is the interaction of the excitations
with the vortex at distances from the vortex line \( r\sim a\ll R \). Therefore,
the corresponding friction force in a trapped condensate can be found in the
local density approximation: We may use Eq.(\ref{Fdef}), derived for a homogeneous
superfluid, and then replace the condensate density \( n_{0} \) by the Thomas-Fermi
density profile of the trapped condensate. 

The Hamiltonian of the single-particle excitations is \( \hbar ^{2}{\bf \hat{k}}^{2}/2m+2n_{0}({\bf r})g-\mu  \),
where the second term originates from the mean-filed interparticle interaction,
with \( n_{0}({\bf r}) \) is the density of the vortex state, \( g=4\pi \hbar ^{2}a_{sc}/m \),
\( a_{sc} \) is the scattering length, and \( \mu =n_{0}(\infty )g \) (\( n_{0}(\infty )\equiv n_{0} \)).
For \( r\rightarrow \infty  \) we have \( \hat{H}({\bf \hat{k}},{\bf r})=\hbar ^{2}{\bf \hat{k}}^{2}/2m+\mu  \).
Hence, the interaction Hamiltonian responsible for the scattering of excitations
from the vortex can be written as
\[
\hat{H}_{int}=2\left[ n_{0}({\bf r})g-\mu \right] .\]
For the vortex charge \( Z=1 \), at distances \( r\ll a \) the interaction
Hamiltonian \( \hat{H}_{int}\approx -2\mu  \). For \( r\gg a \) we have \( n_{0}({\bf r})\approx (\mu -\hbar ^{2}/2mr^{2})/g \),
and \( \hat{H}_{int}\approx -\hbar ^{2}/mr^{2}. \) The scattering amplitude
in Eq.(\ref{Fdef}) can be written as \( f({\bf k},{\bf k}^{\prime })=2\pi \delta (k_{z}-k_{z}^{\prime })\tilde{f}({\bf k},{\bf k}^{\prime }), \)
where the 2D scattering amplitude in the Born approximation is given by 
\begin{equation}
\label{scatampl}
\tilde{f}({\bf k},{\bf k}^{\prime })=\int d^{2}rH_{int}({\bf r})e^{i{\bf qr}}.
\end{equation}
Here \( {\bf q}={\bf k}-{\bf k}^{\prime } \) is the momentum transferred from
the excitation to the vortex. As the amplitude \( \tilde{f} \) only depends
on \( |{\bf q}| \), the first term in Eq.(\ref{Fdef}) is purely longitudinal.

For \( qa\ll 1 \), which corresponds to small angle scattering, from Eq.(\ref{scatampl})
we obtain \( \tilde{f}\sim (\hbar ^{2}/m)\log (1/qa) \). For \( qa\gg 1 \)
we find \( |\tilde{f}(q)|^{2}\sim (\hbar ^{2}/m)^{2}\sin ^{2}(qa-\pi /4)/(aq)^{3}. \)
Using these results in Eq.(\ref{Fdef}), we see that the main contribution to
the integral over momenta comes from energies \( E_{k} \) satisfying the inequality
\( \mu \alt E_{k}\ll T \). A direct calculation of the longitudinal friction
coefficient gives
\begin{equation}
\label{DhighT}
D\approx \kappa \rho _{n}(T)(n_{0}g/T)^{1/2},
\end{equation}
 where the density of the normal component
\[
\rho _{n}=-\frac{1}{3}\int \frac{\partial n}{\partial E_{p}}p^{2}\frac{d^{3}p}{(2\pi \hbar )^{3}}\approx 0.1\frac{m^{5/2}T^{3/2}}{\hbar ^{3}}.\]
A collective character of excitations with energies \( E_{k}\sim \mu  \) can
influence the numerical coefficient in Eq.(\ref{DhighT}), and for this reason
we did not present the exact value of this coefficient in the single-particle
approximation. 

The coefficient \( D\propto T \), and Eq. (\ref{DhighT}) can be rewritten
as \( D\propto \hbar n_{0}\xi  \), where the quantity
\begin{equation}
\label{smallpar}
\xi =(n_{0}a_{sc}^{3})^{1/2}\frac{T}{\mu }\ll 1
\end{equation}
is a small parameter of the finite-temperature perturbation theory at \( T\gg \mu  \).
The inequality \( \xi \ll 1 \) remains valid even near the BEC transition temperature,
except the region of critical fluctuations \cite{fedichev}.

Relying on Eq.(\ref{DhighT}) for the longitudinal friction force, we consider
the motion of the vortex line to the border of the condensate in a static trap,
where the normal component is at rest. Assuming a small friction in Eqs.(\ref{eqofmotion})
and (\ref{forcegen}), for finding a friction-induced small quantity \( {\bf v}_{L}-{\bf v}_{S} \)
we only retain the terms linear in the dissipation coefficients \( D \) and
\( D' \). Then we obtain the equation 
\[
\rho _{s}[({\bf v}_{L}-{\bf v}_{S})\times \mathbold {\kappa }]=-D{\bf v}_{S}-D'[{\bf v}_{S}\times \mathbold {\kappa }]/\kappa \]
which has a solution of the form \( v_{L}=v_{L}^{(r)}\hat{r}+v_{L}^{(\phi )}[\mathbold {\kappa }\times {\bf r}]/\kappa r \).
For the radial (\( v^{(r)} \)) and tangential (\( v^{(\phi )} \)) components
of the velocity of the vortex line we find 
\begin{equation}
\label{cylvel}
\begin{array}{c}
v_{L}^{(r)}=Dv_{S}/\rho _{s}\kappa ,\\
v_{L}^{(\phi )}=v_{S}(1-D'/\rho _{s}\kappa ).
\end{array}
\end{equation}
 From Eqs.(\ref{cylvel}) it is clear that the radial motion of the vortex is
governed by the value of the longitudinal friction coefficient, whereas the
transverse friction (Iordanskii force) simply slows down the drift velocity
(\ref{veldrift}) of the vortex. The radial velocity \( v_{L}^{(r)}\ll v_{S} \),
which is guaranteed by the inequality (\ref{smallpar}).

The time dependence of the distance \( x_{0} \) of the vortex line from the
axis of a cylindrical trap follows from the equation of radial motion for the
vortex, \( dx_{0}/dt=v_{L}^{(r)} \). With Eq.(\ref{cylvel}) for \( v_{L}^{(r)} \)
and Eq.(\ref{veldrift}) for \( v_{S} \), for the characteristic time of motion
of the vortex from the center of the trap to the border we obtain
\begin{equation}
\label{tmotion}
\tau \approx \int _{x_{min}}^{R}\frac{dx_{0}m(R^{2}-x_{0}^{2})\rho _{s}}{\hbar x_{0}\rho _{n}}\left( \frac{n_{0}g}{T}\right) ^{1/2},
\end{equation}
where \( x_{min} \) is the initial displacement of the vortex line from the
axis of the trap. The vortex velocity is the smallest near the trap center,
and the main contribution to the integral in Eq.(\ref{tmotion}) comes from
distances \( x_{0}\ll R \). Therefore, with logarithmic accuracy, we can neglect
\( x_{0} \) in the numerator of the integrand and put \( \rho _{s}=\rho _{S}(0) \),
\( n_{0}=n_{0}(0) \). Then, Eq.(\ref{tmotion}) yields
\begin{equation}
\label{ltfinal}
\tau \approx \frac{mR^{2}\rho _{s}}{\hbar \rho _{n}}\left( \frac{n_{0}g}{T}\right) ^{1/2}\ln (R/x_{min}).
\end{equation}
This result is insensitive to the exact value of \( x_{min} \), and we can
put \( x_{min}\sim a \). 

Once the vortex reaches the border of the condensate, it immediately decays.
Hence, the time \( \tau  \) can be regarded as a characteristic life-time of
the vortex state in a static trap. Interestingly, the decay rate can be written
as
\begin{equation}
\label{rate}
\tau ^{-1}\sim \frac{E_{0}}{\hbar }(n_{0m}a_{sc}^{3})^{1/2}\left( \frac{T}{\mu }\right) ,
\end{equation}
where \( n_{0m} \) is the maximum condensate density, and \( E_{0}\sim \hbar ^{2}/mR^{2} \)
is the energy of excitation corresponding to the motion of the vortex core with
respect to the rest of the condensate (excitation with negative energy, found
in the recent calculations \cite{Oxford,Jap:negative,Bigelow,Rokhsar,Fetter:recent}).
Eq.(\ref{rate}) is similar to the damping rate of low-energy excitations of
a trapped condensate, found beyond the mean-field approach\cite{fedichev,Pitaevskii:review}.
Both rates are proportional to the small parameter \( \xi  \) (\ref{smallpar}).

For \( {\rm Rb} \) and \( {\rm Na} \) condensates at densities \( n_{0}\sim 10^{14}cm^{-3} \)
and temperatures \( 100\alt T\alt 500 \) nK, in the static traps with frequencies
\( 10\alt \omega \alt 100 \) Hz the life-time \( \tau  \) of the vortex state
ranges from \( 0.1 \) to \( 10 \) s. This range of times is relevant for experimental
studies of the dissipative vortex dynamics. 

A proposed way of identifying the presence of a vortex state in a trapped Bose-condensed
gas assumes switching off the trap and observing a ballistically expanding gas
sample\cite{YC:last}. As follows from the numerical simulations \cite{YC:last},
at zero temperature the expansion of a condensate with a vortex occurs along
the lines of the scaling theory \cite{gora:scaling,YC:scaling}. The shape of
the Bose-condensed state is nearly preserved and its spatial size is increasing.
Due to expansion the density of the condensate decreases, and the size of the
vortex core increases to match the instantaneous value of the healing length.
This should allow one to detect the vortex through the observation of a hole
in the density profile of the condensate. 

It is important to emphasize that at temperatures \( T\gg \mu  \) the thermal
cloud will expand with the thermal velocity \( v_{T}\sim \sqrt{T/m} \) which
is much larger than the expansion velocity of the condensate (the latter is
of order the sound velocity \( c_{s} \)). Therefore, after a short time \( R/v_{T} \)
the thermal component flies away, and the dissipation-induced motion of the
vortex core ceases. Accordingly, the expansion of the Bose-condensed state will
be essentially the same as that at zero temperature. This means that the relative
displacement of the vortex core from the trap center practically remains the
same as before switching off the trap. Therefore, the dissipative dynamics of
the vortex state in the initial static trap, i.e. the motion of the vortex core
towards the border of the condensate, can be studied by switching off the trap
at different times and visualizing the position of the vortex core in a ballistically
expanding condensate.

In conclusion, we have developed a theory of dissipative dynamics of a vortex
state in a trapped Bose-condensed gas at finite temperatures and calculated
the decay time of the vortex with charge equal to \( 1 \) in a static trap.
Our theory can be further developed to analyze the motion of vortices in rotating
traps and, in particular, to calculate a characteristic time of the formation
of the vortex state in a trap rotating with supercritical frequency.

This work was supported by the Stichting voor Fundamenteel Onderzoek der Materie
(FOM), by INTAS, and by the Russian Foundation for Basic Studies. We acknowledge
fruitful discussions with J. Dalibard, J.T.M. Walraven, D.S. Petrov and A.E.
Muryshev. 

\( ^{*} \)L.K.B. is an unit\'e de recherche de l'Ecole Normale Sup\'eriere
et de l'Universit\'e Pierre et Marie Curie, associ\'ee au CNRS.

\bibliographystyle{prsty}
\bibliography{vortex}

\begin{thebibliography}{10}

\bibitem{Cor95}
M.~H. Anderson {\it et~al.}, Science {\bf 269},  198  (1995).

\bibitem{Ket95}
K.~B. Davis {\it et~al.}, Phys. Rev. Lett. {\bf 75},  3969  (1995).

\bibitem{Hul95}
C.~C. Bradley, C.~A. Sackett, J.~J. Tolett, and R.~G. Hulet, Phys. Rev. Lett.
  {\bf 75},  1687  (1995).

\bibitem{Pitaevskii:review}
F. Dalfovo, S. Giorgini, L. Pitaevskii, and S. Stringari, Rev. Mod. Phys. {\bf
  71},  463  (1999).

\bibitem{Legget}
A. Leggett,  in {\em Topics in superfluidity and superconductivity}, {\em Low
  Temperature Physics}, edited by M. Hoch and R. Lemmer (Springer Verlag, New
  York, 1992).

\bibitem{Stringari:vortices2}
S. Stringari, Phys. Rev. Lett. {\bf 82},  4371  (1999).

\bibitem{Zurek}
J.~R. Anglin and W.~H. Zurek, e-print quant-ph/9804035.

\bibitem{Zoller:vorticesprod}
R. Dum, I. Cirac, M. Lewenstein, and P. Zoller, Phys. Rev. Lett. {\bf 80},
  2972  (1998).

\bibitem{olshanii:vortices}
M. Olshanii and M. Naraschewski, e-print cond-mat/9811314.

\bibitem{propTurb}
B. Jackson, J.~F. McCann, and C.~S. Adams, Phys. Rev. Lett. {\bf 80},  3903
  (1998).

\bibitem{YC:last}
Y. Castin and R. Dum, Europ. Phys. Journal D  (1999).

\bibitem{anglin:kink}
T. Busch and J. Anglin, e-print cond-mat/9809408.

\bibitem{gora:kink}
A.~E. Muryshev, H.~B. van Linden~van Heuvell, and G.~V. Shlyapnikov, e-print
  cond-mat/9811408.

\bibitem{Rokhsar}
D.~S. Rokhsar, Phys. Rev. Lett. {\bf 79},  2161  (1997).

\bibitem{Fetter:dissipation}
A.~L. Fetter, J. Low. Temp. Phys. {\bf 113},  189  (1998).

\bibitem{Fetter:recent}
A. Svidzinsky and A. Fetter, e-print cond-mat/9811348.

\bibitem{Bigelow}
H. Pu, C.~K. Law, J.~H. Eberly, and N. Bigelow, Phys. Rev. A {\bf 59},  1533
  (1999).

\bibitem{Donelly}
R.~J. Donnelly, {\em Quantized vortices in helium II} (Cambridge University
  Press, Cambridge, 1991).

\bibitem{Sonin}
E.~B. Sonin, Phys. Rev. B {\bf 55},  485  (1997).

\bibitem{PAo:vortex}
D.~J. Thouless, P. Ao, and Q. Niu, Phys. Rev. Lett. {\bf 76},  3758  (1996).

\bibitem{Onsager}
L. Onsager, Nouvo Cimento {\bf 6},  249  (1949).

\bibitem{SP2}
E.~M. Lifshitz and L.~P. Pitaevskii, {\em Statistical Physics, Part 2}
  (Pergamon Press, Oxford, 1980).

\bibitem{Pitaevskii:friction}
L.~P. Pitaevskii, Sov. Phys.-JETP {\bf 8},  888  (1959).

\bibitem{Fetter:friction}
A.~L. Fetter, Phys. Rev. {\bf 136},  A1488  (1964).

\bibitem{Iordanskii}
S.~V. Iordanskii, Sov. Phys.-JETP {\bf 22},  160  (1966).

\bibitem{fedichev}
P.O. Fedichev, G.V. Shlyapnikov, and J.T.M. Walraven, Phys. Rev. Lett. {\bf
  80}, 2269 (1998); P.O. Fedichev and G.V. Shlyapnikov, Phys. Rev. A {\bf 58},
  3146 (1998).

\bibitem{Oxford}
R.~J. Dodd, K. Burnett, M. Edwards, and C. Clark, Phys. Rev. A {\bf 56},  587
  (1997).

\bibitem{Jap:negative}
T. Isoshima and K. Machida, J. Phys. Soc. Jpn. {\bf 66},  3602  (1997).

\bibitem{gora:scaling}
Y. Kagan, E.~L. Surkov, and G.~V. Shlyapnikov, Phys. Rev. A {\bf 54},  R1753
  (1996), ibid {\bf 55}, R18 (1997).

\bibitem{YC:scaling}
Y. Castin and R. Dum, Phys. Rev. Lett. {\bf 77},  5315  (1996).

\end{thebibliography}

\end{document}